\documentstyle[12pt]{article}
\vspace{1.8cm} \oddsidemargin 0pt \evensidemargin 0pt  \topmargin
-1.5cm \footheight 26pt \footskip 10mm

\begin{document}
\baselineskip 20pt
\begin{center}
\baselineskip=24pt {\Large  Nonlinear Quantum Wave Equation of
Radiation Electron and Dissipative Systems}

\vspace{1cm} {Xiang-Yao Wu$^{a}$ \footnote{E-mail:wuxy2066@163.com
}, Xiao-Jing Liu$^{a}$, Bai-Jun Zhang$^{a}$ and Yi-Heng Wu$^{a}$
 } \vskip 10pt \noindent{\footnotesize a.
\textit{Institute of Physics, Jilin Normal University, Siping
136000, China}}

\end{center}
\date{}
\renewcommand{\thesection}{Sec. \Roman{section}} \topmargin 10pt
\renewcommand{\thesubsection}{ \arabic{subsection}} \topmargin 10pt
{\vskip 5mm
\begin {minipage}{140mm}
\centerline {\bf Abstract} \vskip 8pt
\par
\indent\\

\hspace{0.3in} As well known, an electron will produce radiative
reaction force when the electron emits electromagnetism radiation.
The electron radiative effect had not been considered in
Schr\"{o}dinger wave equation. In this paper, we give the
nonlinear quantum wave equations
for the radiative electron and some dissipative systems.


\end {minipage}

\newpage
\section * {1. Introduction }

\hspace{0.3in}Linearity of the Schr\"{o}dinger equation and the
validity of the superposition principle have been essential
ingredients of quantum theory since its earliest days. Practically
all physical phenomena behave nonlinearly when examined over a
sufficiently large range of the dynamical parameters that
determine their evolution.

A number of earlier works that have attempted to extend quantum
theory in a nonlinear way, there are: The work by Kibble and
Randjbar-Daemi is close to ours in that they consider how
nonlinear modifications of quantum field theory can be made
compatible with Lorentz or more generally coordinate invariance
[1, 2]. Besides considering a coupling of quantum fields to
classical gravity according to general relativity, which induces
an intrinsic nonlinearity [2, 3], these authors study mean-field
type nonlinearities, where parameters of the model are state
dependent through their assumed dependence on expectations of
certain operators. Work by Bialynicki-Birula and Mycielski
introduces a logarithmic nonlinearity into the nonrelativistic
Schr\"{o}dinger equation, with which many of the features of
standard quantum mechanics are left intact [4]. A number of
different nonrelativistic models of this kind have been
systematically studied by Weinberg, offering also an assessment of
the observational limits on such modifications of the
Schr\"{o}dinger equation [5]. Independently, Doebner and Goldin
and collaborators have also studied nonlinear modifications of the
nonrelativistic Schr\"{o}dinger equation [6]. This was originally
motivated by attempts to incorporate dissipative effects. Later,
however, they have shown that classes of nonlinear Schr\"{o}dinger
equations, including many of those considered earlier, can be
obtained through nonlinear transformations of the linear quantum
mechanical equation. The nonlinear quantum mechanics has a
practical importance in different fields, like condensed matter,
quantum optics and atomic and molecular physics; even quantum
gravity may involve nonlinear quantum mechanics [7-10]. Another
important example is in the modern field of quantum computing
[11-14].

As well known, an electron will produce radiative reaction force
when the electron emits electromagnetism radiation. The electron
radiative effect is not considered in Schr\"{o}dinger wave
equation and Dirac quantum theory. In this paper, we study the
quantum effect of lower and higher energy electron radiative
reaction, and give the nonlinear quantum wave equation to describe
the nonrelativistic and relativistic radiation electron.

\section * {2. Nonlinear quantum wave equation for nonrelativistic radiation electron}

\hspace{0.3in}The radiation reaction force of low energy electron
can argument based on conservation of energy for a nonrelativistic
electron. An electron of mass $m$ and charge $e$ acted on by an
external force $\vec{F}_{ext}$ moves according to the Newton
equation of motion: $ m\dot{\vec{v}}=\vec{F}_{ext}$. To account
for this radiative energy loss and its effect on the motion of the
electron we should modify the Newton equation by adding a
radiative force $\vec{F}_{rad}$, and the Newton equation is
$m\dot{\vec{v}}=\vec{F}_{ext}+\vec{F}_{rad}$.

Since the electron is accelerated, it emits radiation at a rate
given by Larmor power formula [15]
\begin{equation}
p=\frac{1}{4\pi\varepsilon_{0}}\frac{2e^{2}}{3c^{3}}
|\dot{\vec{v}}|^{2}, \hspace{0.3in}   (v\ll c)
\end{equation}
The electron radioactive energy in unit time is
\begin{equation}
\frac{dw}{dt}=\frac{1}{4\pi\varepsilon_{0}}\frac{2e^{2}}{3c^{3}}
|\dot{\vec{v}}|^{2},
\end{equation}
and the radioactive energy in the time interval $(0, t)$ is
\begin{equation}
W=\frac{1}{4\pi\varepsilon_{0}}\int_{0}^{t}\frac{2e^{2}}{3c^{3}}
|\dot{\vec{v}}|^{2}dt.
\end{equation}
By conservation of energy, we have
\begin{equation}
\frac{d}{dt}(T+V)=-\frac{d}{dt}W,
\end{equation}
or
\begin{equation}
E=T+V+W=constant,
\end{equation}
we directly promote all classical variables $E$, $T$ and $W$ to
quantum operators
\begin{equation}
\hat{E}\rightarrow i\hbar\frac{\partial}{\partial
t},\hspace{0.3in} \hat{p}\rightarrow -i\hbar\nabla,
\end{equation}
Eq. (5) becomes
\begin{equation}
i\hbar\frac{\partial}{\partial
t}=-\frac{{\hbar}^{2}}{2m}{\nabla}^{2}+V+
\frac{1}{4\pi\varepsilon_{0}}\frac{2e^{2}}{3c^{3}}\int_{0}^{t}
\dot{\vec{v}}^{2}dt',
\end{equation}
and act the wave function $\psi(\vec{r},t)$ on the right side of
Eq. (7), we have
\begin{equation}
i\hbar\frac{\partial}{\partial
t}\psi(\vec{r},t)=-\frac{{\hbar}^{2}}{2m}{\nabla}^{2}\psi(\vec{r},t)+V\psi(\vec{r},t)+
\frac{1}{4\pi\varepsilon_{0}}\frac{2e^{2}}{3c^{3}}\int_{0}^{t}
\dot{\vec{v}}^{2}dt'\cdot\psi(\vec{r},t),
\end{equation}
The complex time dependent wave function $\psi(\vec{r},t)$ is
expressed in terms of a Lagrange action $S(\vec{r},t)$ [8]
\begin{equation}
\psi(\vec{r},t)=\sqrt{\rho(\vec{r},t)}e^{iS(\vec{r},t)/\hbar},
\end{equation}
and
\begin{equation}
\frac{\psi(\vec{r},t)}{\psi^{*}(\vec{r},t)}=e^{2iS(\vec{r},t)/\hbar}.
\end{equation}
Using the relation from the Hamilton-Jacobi theory
\begin{equation}
\vec{p}=\nabla S(\vec{r},t)=m\vec{v},
\end{equation}
or
\begin{equation}
\vec{v}=\frac{\nabla S(\vec{r},t)}{m}.
\end{equation}
From Eqs. (10)-(12), we can express the velocity and acceleration
in terms of $\psi(\vec{r},t)$ and $\psi^{*}(\vec{r},t)$ as follows
\begin{equation}
\vec{v}=-i\frac{\hbar}{2m}\nabla
\ln\frac{\psi(\vec{r},t)}{\psi^{*}(\vec{r},t)},
\end{equation}
and
\begin{equation}
\dot{\vec{v}}=-i\frac{\hbar}{2m}\frac{\partial}{\partial t}\nabla
\ln\frac{\psi(\vec{r},t)}{\psi^{*}(\vec{r},t)},
\end{equation}
and then Eq. (8) becomes
\begin{equation}
i\hbar\frac{\partial}{\partial
t}\psi(\vec{r},t)=-\frac{{\hbar}^{2}}{2m}{\nabla}^{2}\psi(\vec{r},t)+V\psi(\vec{r},t)-
\frac{1}{4\pi\varepsilon_{0}}\frac{2e^{2}}{3c^{3}}\frac{\hbar^{2}}{4m^{2}}\int_{0}^{t}
(\frac{\partial}{\partial t'}\nabla
\ln\frac{\psi(\vec{r},t')}{\psi^{*}(\vec{r},t')})^{2}dt'\cdot\psi(\vec{r},t).
\end{equation}
The Eq. (15) is radioactive electron nonlinear quantum wave
function for the low energy radioactive electron.

\section * {3. Nonlinear quantum wave equation for the dissipative systems}
\hspace{0.3in}The heavy ion scattering experiments give strong
indication for a reaction called deep inelastic process [8]. The
heavy ion lose their entire available kinetic energy during the
collision and are then repelled just by their mutual Coulomb
interaction energy. Thus nuclear friction seems to play an
important role. Moreover, there is also evidence that the fission
process is damped during the descent from saddle to scission.
Classical as well as microscopic calculations in these phenomena
which include frictional effects have already been made [9, 10,
11]. Quantal friction , however, is still an open problem. Other
applications are, for instance, the motion of Stokes' ball in a
viscous medium, Brownian motion, or an electric oscillator
composed of inductance, capacitor, and resistor. In the following,
we should give the quantum wave equation for these dissipative
systems.

1. The dissipative system for the dissipative force
$\vec{F_{n}}=-k\vec{v}$. \\
For a dissipative system, there is a conserved quantity, which is
the sum of a particle's mechanical energy and the work doing by
the dissipative force acting on the particle. It is
\begin{equation}
E=T+V-\int \vec{F}_{n} \cdot d\vec{r}=constant,
\end{equation}
where $T$ and $V$ are the particle's kinetic energy and potential
energy, respectively, the integrate $\int \vec{F}_{n} \cdot
d\vec{r}$ is the work doing by the dissipative force acting on the
particle, and $E$ is the system's total energy. \\
Substituting $\vec{F}=-k\vec{v}$ into Eq. (16), we have
\begin{equation}
E=T+V+k\int_{0}^{t} \vec{v}^{2} dt',
\end{equation}
substituting Eqs. (6), (13) into (17), we obtain the operator
equation
\begin{equation}
i\hbar\frac{\partial}{\partial
t}=-\frac{\hbar^{2}}{2m}\nabla^{2}+V+k\int_{0}^{t}-\frac{\hbar^{2}}{4m^{2}}
(\nabla \ln\frac{\psi(\vec{r},t')}{\psi^{*}(\vec{r},t')})^{2}dt',
\end{equation}
and act the wave function $\psi(\vec{r}, t)$ on the right side of
Eq. (18), we have
\begin{equation}
i\hbar\frac{\partial}{\partial t}\psi(\vec{r},
t)=-\frac{\hbar^{2}}{2m}\nabla^{2}\psi(\vec{r}, t)+V\psi(\vec{r},
t)-\frac{\hbar^{2}k}{4m^{2}}\int_{0}^{t} (\nabla
\ln\frac{\psi(\vec{r},t')}{\psi^{*}(\vec{r},t')})^{2}dt'\cdot\psi(\vec{r},
t),
\end{equation}
The Eq. (19) is the nonlinear quantum wave equation corresponding
to the dissipative force $\vec{F}=-k\vec{v}$.

2. The dissipative system for the dissipative force
$\vec{F_{n}}=-k\vec{v}^{2}\frac{\vec{v}}{|\vec{v}|}$.\\
Substituting $\vec{F_{n}}=-k\vec{v}^{2}\frac{\vec{v}}{|\vec{v}|}$
into Eq. (16), we have
\begin{equation}
E=T+V+k\int_{0}^{t} v^{3}dt'.
\end{equation}
Similarly, we can obtain the nonlinear quantum wave equation
corresponding to the dissipative force
$\vec{F}=-k\vec{v}^{2}\frac{\vec{v}}{|\vec{v}|}$. It is
\begin{equation}
i\hbar\frac{\partial}{\partial t}\psi(\vec{r},
t)=-\frac{\hbar^{2}}{2m}\nabla^{2}\psi(\vec{r}, t)+V\psi(\vec{r},
t)+ik\frac{\hbar^{3}}{8m^{3}}\int_{0}^{t} (\nabla
\ln\frac{\psi(\vec{r},t')}{\psi^{*}(\vec{r},t')})^{3}dt'\cdot\psi(\vec{r},
t).
\end{equation}

3. The dissipative system for the dissipative force
$\vec{F}=-k\dot{\vec{v}}$.\\
Substituting $\vec{F}=-k\dot{\vec{v}}$ into Eq. (16), we have
\begin{equation}
E=T+V+k\int_{0}^{t}\dot{\vec{v}}\cdot\vec{v} dt'.
\end{equation}
We can obtain the nonlinear quantum wave equation corresponding to
the dissipative force $\vec{F}=-k\dot{\vec{v}}$. It is
\begin{equation}
i\hbar\frac{\partial}{\partial t}\psi(\vec{r},
t)=-\frac{\hbar^{2}}{2m}\nabla^{2}\psi(\vec{r}, t)+V\psi(\vec{r},
t)-\frac{\hbar^{2}}{4m^{2}}k\int_{0}^{t} \nabla
\ln\frac{\psi(\vec{r},t')}{\psi^{*}(\vec{r},t')}\cdot\nabla
\frac{\partial}{\partial
t'}\ln\frac{\psi(\vec{r},t')}{\psi^{*}(\vec{r},t')}dt'\cdot\psi(\vec{r},t).
\end{equation}
\section * {4. Conclusion}
The electron in an atom produces radiative reaction. We should
consider the quantum effect of the radiation when we calculate the
atom energy spectrum and wave functions. The nonlinear term in Eq.
(15) is the radiation quantum correction. For the dissipative
systems, we can research their quantum property by the new
nonlinear quantum wave equations.

\end{document}